\documentclass[runningheads]{llncs}
\usepackage[T1]{fontenc}

\usepackage{amsmath}
\usepackage{enumitem}
\usepackage{cleveref}

\usepackage{float}
\usepackage{listings}
\usepackage{xcolor}

\definecolor{codegreen}{rgb}{0,0.6,0}
\definecolor{codegray}{rgb}{0.5,0.5,0.5}
\definecolor{codepurple}{rgb}{0.58,0,0.82}
\definecolor{backcolour}{rgb}{0.95,0.95,0.92}

\lstdefinestyle{mystyle}{
    backgroundcolor=\color{backcolour},   
    commentstyle=\color{codegreen},
    keywordstyle=\color{magenta},
    numberstyle=\tiny\color{codegray},
    stringstyle=\color{codepurple},
    basicstyle=\ttfamily\footnotesize,
    breakatwhitespace=false,         
    breaklines=true,                 
    captionpos=b,                    
    keepspaces=true,                 
    numbers=left,                    
    numbersep=5pt,                  
    showspaces=false,                
    showstringspaces=false,
    showtabs=false,                  
    tabsize=2
}

\newfloat{code}{htbp}{loc}
\floatname{code}{Listing}

\crefname{code}{Listing}{Listings}
\Crefname{code}{Listing}{Listings}   

\usepackage{fancyvrb}

\begin{document}
\title{Ancient Algorithms for a Modern Curriculum}
%
%
\author{Aalok Thakkar}
\authorrunning{A. Thakkar}
%
\institute{Ashoka University}
\maketitle              
\begin{abstract}
Despite growing calls for inclusive computing education, algorithm instruction remains largely decontextualized and Eurocentric. We present a pedagogical approach that integrates ancient algorithms from the India into foundational CS courses. Through a mixed-methods study involving 166 students across four sections, we demonstrate that historically grounded algorithm instruction increases student engagement, improves algorithmic thinking skills, and promotes more inclusive perspectives on computing. Our findings suggest that contextualizing algorithms within their cultural and historical origins can bridge the gap between familiar mathematical procedures and formal CS concepts while addressing representation gaps in computing education.
\end{abstract}

\section{Area and Context}
Despite ongoing calls for inclusive and culturally responsive pedagogy in computing education \cite{Margolis2002,Scott2015}, the teaching of algorithms remains largely decontextualized. Foundational computer science courses often present algorithmic thinking as purely formal and ahistorical, emphasizing efficiency, correctness, and abstraction. When history is mentioned, it usually centers on the modern development of digital computers, highlighting figures such as Turing, von Neumann, and Babbage. This narrow view misrepresents the origins of algorithmic reasoning and perpetuates a Eurocentric worldview that undermines equity and representation in STEM \cite{Benjamin2019,Dignazio2020}. In contrast, algorithmic thinking predates electronic computers by millennia and has deep roots in ancient civilizations including India, China, Babylon, and Egypt \cite{Ifrah2001,Knuth1997}. Our work responds to this gap by embedding algorithm instruction in broader historical and cultural contexts, with particular attention to classical Indian contributions.
\section{Best Practice}
\label{sec:best_practice}

In two courses FC-0306 Quantitative Reasoning and Mathematical Thinking, and CS-1102 Introduction to Computer Science at Ashoka University, we adopted a historically grounded, cross-cultural approach to teaching algorithmic thinking. Rather than presenting algorithms solely as abstract procedures, we framed them as \textit{cultural artefacts} that are embedded in mathematical, poetic, and problem-solving traditions across time and place. This approach foregrounds the human story behind algorithmic thinking, engaging students by linking computational ideas to their historical and cultural origins. Our method aligns with constructivist and culturally responsive pedagogies, while providing students with concrete experiences of algorithmic generality, recursion, and abstraction.

\subsection{Right-Angle Reasoning and the Pothayanar Rule}
\label{subsec:triangle}

We introduced the topic of right triangles through an example drawn from Tamil mathematics: a rule attributed to the poet-mathematician \textit{Pothayanar} (sometimes rendered “Podhayanar”) that estimates the hypotenuse $c$ of a right triangle with legs $a$ and $b$ as
\begin{equation*}
c = \frac{7}{8}a + \frac{1}{2}b.
\end{equation*}

This rule yields accurate results for certain familiar triples such as $(3,4,5)$, $(5,12,13)$, and their scaled variants, but fails for arbitrary inputs. As students test the rule computationally, they quickly notice its empirical validity only in select cases. This leads naturally into a discussion of verification versus proof: as Dijkstra famously observed, “Testing can show the presence of bugs, not their absence.” 

We contrast Pothayanar’s empirical rule with a general constructive identity from the \textit{Katyayana Śulbasūtra} (c.~8th~century~BCE):
\begin{equation*}
(m^2 - n^2)^2 + (2mn)^2 = (m^2 + n^2)^2,
\end{equation*}
which generates all primitive Pythagorean triples. By juxtaposing these two approaches, students learn to distinguish between an \textit{algorithmic pattern} that works by observation and one grounded in a generative proof schema. The exercise naturally motivates discussion of abstraction, correctness, and generality, which are core values in computer science education.

\subsection{Recursive Decomposition and Piṅgala’s Binary Prosody}
\label{subsec:pingala}

To reinforce recursive thinking, we draw upon the \textit{Chandaḥśāstra} (c.~3rd~century~BCE), a foundational Sanskrit treatise on poetic meter by Piṅgala. Piṅgala represented syllables as either long (\textit{guru}) or short (\textit{laghu}), effectively introducing a binary encoding of metrical patterns. Each line of poetry may thus be viewed as a binary string, with recursive generation rules that anticipate modern binary enumeration.

Piṅgala’s description of recursive doubling mirrors the divide-and-conquer principle underlying \textit{fast exponentiation}. 
In modern notation, the recursive exponentiation algorithm can be expressed as

\begin{equation*}
\mathrm{exp}_2(n) =
\begin{cases}
1, & \text{if } n = 0;\\[6pt]
\mathrm{exp}_2\!\left(\tfrac{n}{2}\right)^{2}, & \text{if $n$ is even;}\\[6pt]
2 \times \mathrm{exp}_2\!\left(\tfrac{n-1}{2}\right)^{2}, & \text{if $n$ is odd.}
\end{cases}
\end{equation*}

Here, $\mathrm{exp}_2(n)$ computes $2^n$ using recursive halving of the exponent. 
In the classroom, students are first guided to trace the recursion tree by hand for small values of $n$, 
observing how each step halves the exponent until reaching the base case. 
They then generalize the computation from $\mathrm{exp}_2(n)$ to $\mathrm{exp}(x, n)$, yielding a recursive routine that can compute $x^n$ for any integer base~$x$.
This generalization reinforces abstraction.
By comparing this algorithmic schema to Piṅgala’s metrical doubling, students recognize a shared structural pattern. 
In both cases, recursive decomposition produces exponential growth of outcomes from logarithmic depth of computation.

Piṅgala’s enumeration of metrical forms for verses of length~$n$ follows the well-known Fibonacci recurrence:
\begin{equation*}
F_n = F_{n-1} + F_{n-2},
\end{equation*}
where $F_n$ counts the total number of permissible metrical patterns of length~$n$. 
This relationship arises from the task of making meters of a fixed cadence. A meter of cadence $n$
either begins with a short (\textit{laghu}) followed by a pattern of length~$n-1$, 
or begins with a long (\textit{guru}), which occupies two cadence units and is followed by a pattern of length~$n-2$. 

Piṅgala’s \textit{Meru Prastāra} (“mountain of arrangements”) offers a recursive structure, 
in which each entry is obtained as the sum of the two directly above it, thereby defining the binomial coefficients:
\begin{equation*}
\binom{n}{k} = \binom{n-1}{k-1} + \binom{n-1}{k}.
\end{equation*}
In our course, this construction serves as a bridge between poetic enumeration and combinatorial reasoning.  
Through guided discussion, they connect the \textit{Meru Prastāra} to the later work of Blaise~Pascal and Pierre~de~Fermat on the seventeenth-century ``problem of points,'' 
in which the stakes of an interrupted game are divided according to the probabilities of each player’s eventual victory. 

This comparative framing allows students to see algorithmic ideas as historically grounded. They gain an appreciation for recursion not only as a programming technique but as a universal mode of reasoning about structure, probability, and combinatorial growth. 

\subsection{Digit-by-Digit Root Extraction: Aryabhata’s Algorithm}
\label{subsec:aryabhata}

The third study examines the root-extraction algorithm described by \textit{Āryabhaṭa} (476~CE) in the \textit{Āryabhaṭīya}. 
His digit-by-digit square-root method, grounded in the decimal place-value system, constructs the root incrementally while maintaining a running remainder. 
Students first execute the procedure manually for small examples (e.g., $\sqrt{1521}$), observing how each step determines a new digit and updates the residual value. 
They then implement the algorithm in OCaml, using recursion and explicit loop invariants to capture the evolving relationship between divisor, quotient digit, and remainder. 
We contrast this approach with approximation-based techniques such as Heron’s method and the iteration in the Bakhshālī manuscript, prompting discussion about exact versus iterative computation and the algorithmic notion of convergence.

\subsection{Solving Indeterminate Equations: Kuttaka and Chakravala}
\label{subsec:kuttaka}

To deepen understanding of recursion and iteration, we next introduced classical Indian methods for solving indeterminate equations, notably the \textit{Kuttaka} (``pulveriser'') and the \textit{Chakravala} (``cyclic'') algorithms. The Kuttaka method, described by Aryabhata, solves linear Diophantine equations of the form $
a x + b y = c$,
using a recursive process similar. 
The Chakravala method, elaborated by Bhāskara II (12th c.), iteratively refines triples $(a, b, k)$ to solve Pell-type equations of the form $
x^2 - N y^2 = 1$.
This provides a context for exploring algorithmic convergence, optimization of parameters, and state-space exploration. Students implemented a simplified Chakravala loop, observing how the algorithm ``cycles'' through candidate triples until $k=1$ is achieved.

Framing these techniques as puzzles resonated strongly with students. Their implementations made explicit use of recursion and modular arithmetic, connecting naturally to topics such as dynamic programming and iterative refinement.

\subsection{Global and Comparative Algorithmic Traditions}
\label{subsec:global}

Finally, to broaden cultural scope, we incorporated algorithmic examples from Egyptian fractions, Heron’s method, Sieve of Eratosthenes, and Euclid’s proof of the infinitude of primes.
These comparative examples highlight that algorithmic design is a \textit{shared human endeavour}, transcending geography and time. They also make visible the iterative evolution of computational ideas, reinforcing to students that algorithms are part of a long intellectual continuum rather than a purely modern construct.

\section{Justification}
This approach is innovative because it resists the framing of algorithms as culturally neutral mechanical abstractions. By embedding algorithm instruction in cultural and historical context, we reveal algorithms as products of human creativity, shaped by poetry, ritual, and problem-solving traditions. Such reframing not only enriches student understanding of computation but also fosters inclusivity by highlighting non-Western contributions. Students come to see computing as a diverse, human-centered practice rather than a narrow technical domain.

\section{Insights}
The approach yielded several benefits: students showed increased engagement, deeper appreciation of global contributions to computing, and a stronger connection between abstract algorithmic ideas and their own cultural identities. At the same time, challenges emerged. Integrating historical material into a tightly packed syllabus required significant trade-offs. Reliable translations and interpretations of ancient texts were not always easy to source. Assessment design also proved difficult, as traditional exams did not align well with the exploratory and reflective goals of the pedagogy. Nonetheless, framing algorithms as puzzles and connecting them to cultural contexts consistently improved motivation and comprehension.

\section{Conclusion}
Our historically grounded, cross-cultural pedagogy reframes algorithm instruction as a human endeavor spanning civilizations and generations. This method not only deepens student engagement but also disrupts Eurocentric narratives by foregrounding non-Western contributions. Although implementation requires careful planning, the positive reception suggests that this approach has strong potential to enrich computing education and support inclusivity.

\section{Suggestions for Others}
Educators interested in adopting this approach may wish to begin with a small set of well-documented historical examples and gradually expand. Developing open-access repositories of code examples, translations, and classroom activities would ease adoption and encourage adaptation. Experimenting with diverse assessment formats, such as projects, creative writing, or peer teaching, can better align evaluation with exploratory goals. Future work might track long-term impacts on student identity and persistence, and extend the cross-cultural lens to other areas such as discrete mathematics, data structures, and design labs. Potential pitfalls include overloading the syllabus, relying on unreliable translations, or treating historical material superficially. With thoughtful integration, however, this method provides a powerful path toward more inclusive and engaging computing education.

\end{document}